\documentclass[%
 reprint,
 amsmath,amssymb,
 aps,
]{revtex4-2}
\usepackage{mathrsfs}
\usepackage{color}
\usepackage{ulem}
\usepackage{graphicx}
\usepackage{dcolumn}
\usepackage{bm}

\begin{document}

\preprint{APS/123-QED}

\title{Effect of Coulomb Interaction on Seebeck Coefficient of Organic Dirac Electron System $\alpha$-(BEDT-TTF)$_2$I$_3$}

\author{D. Ohki$^1$}
\email{dohki@s.phys.nagoya-u.ac.jp}
\author{Y. Omori$^2$}%
\author{A. Kobayashi$^1$}
 \affiliation{$^1$Department of Physics, Nagoya University, Furo-cho, Chikusa-ku, Nagoya, 464-8602 Japan \\
  $^2$Toyota College, National Institute of Technology, Eisei-cho 2-1, Toyota, 471-8525 Japan \\
}%




\date{\today}


\begin{abstract}
Motivated by the results of recent thermoelectric effect studies, we show the effects of Coulomb interactions on the Seebeck coefficient based on an extended Hubbard model that describes the electronic states of a slightly doped organic Dirac electron system, $\alpha$-(BEDT-TTF)$_2$I$_3$. 
Our results indicate that the Hartree terms of the Coulomb interactions enhance the electron-hole asymmetry of the energy band structure and change the energy dependence of the relaxation time from impurity scattering, which reflects the shape of the density of states. 
Thus, the Seebeck coefficient exhibits a non-monotonic $T$ dependence which qualitatively agrees with the experimental results. 
Furthermore, we also show that the signs of the Seebeck coefficient and the Hall coefficient calculated by linear response theory do not necessarily correspond to the sign of the chemical potential using a modified Weyl model with electron-hole asymmetry. 
These results point out that changing the electron-hole asymmetry by strong Coulomb interaction has the potential to controllable the sign and value of the Seebeck coefficient in the Dirac electron systems. 
\end{abstract}

\maketitle


\section{\label{sec:level1}Introduction}

%
The organic conductor $\alpha$-(BEDT-TTF)$_2$I$_3$ has a two-dimensional (2D) massless Dirac electron (DE) system in the high pressure region~\cite{Kajita1992, Tajima2000, Kobayashi2004, Katayama2006, Kobayashi2007, Goerbig2008, Kajita2014}. It shows unique transport properties, such as the inter-band effect of the magnetic field in the Hall effect~\cite{Kobayashi2008,Tajima2012} and the giant Nernst effect~\cite{Proskurin2013,Konoike}.
By contrast,  in a low temperature and low pressure region, a charge-ordering (CO) insulator phase appears, where the mass of the DE is induced by breaking the inversion symmetry~\cite{Bender1984,KinoFukuyama, Seo2000, TakahashiStripe, Kakiuchi2007}. 
The transition temperature is $T_{\rm CO}=135$ K  at ambient pressure, and it decreases linearly as the hydrostatic pressure $P$ increases and becomes zero at $P=P_{\rm C}\simeq 12$ kbar. 

The electron correlation effects play important roles in both phases.
The CO phase is induced by nearest-neighbor Coulomb interactions~\cite{KinoFukuyama, Seo2000,Seo2004,Kobayashi2011PRB} and exhibits anomalous properties on the spin gap~\cite{Tanaka2016, Ishikawa2016} and transport phenomena in $\alpha$-(BEDT-TTF)$_2$I$_3$~\cite{Liu, Beyer, Matsuno2016, Omori2017,Ohki2018JPSJ,Ohki2018Crystals,Ohki2019}. 
In the massless DE phase, the long-range Coulomb interaction suppresses the magnetic susceptibility, owing to Dirac cone reshaping and ferromagnetic polarization~\cite{Hirata2016, Matsuno2017, Matsuno2018}, and
it enhances spin-triplet excitonic fluctuations, owing to perfect electron-hole nesting under an in-plane magnetic field~\cite{Hirata2017}.

The thermoelectric performance of materials is often characterized by a Seebeck coefficient, which defined as the electromotive force induced by a temperature gradient. 
It is suggested in recent years that the electron correlation effects also makes an important contribution to a thermoelectric effect. 
For instance, a giant Seebeck coefficient in low temperature caused by the electron correlation effect is reported in organic compounds such as (TMTSF)$_2$PF$_6$~\cite{Machida}. 
Thus, $\alpha$-(BEDT-TTF)$_2$I$_3$ is also expected as strongly correlated thermoelectric material, and attracts attention in both theoretical and experimental aspects. 

Recently, an anomalous Seebeck effect was observed in $\alpha$-(BEDT-TTF)$_2$I$_3$~\cite{Kitamura, Konoike}. 
The Seebeck coefficient in the massless DE phase shows a positive value.
It forms a gentle peak at approximately 50 K, and decreases linearly toward absolute zero as $T$ decreases under high pressure $P > P_{\rm C}$.
Under low pressure $P < P_{\rm C}$, the Seebeck coefficient exhibits a sharp positive peak at approximately $T_{\rm CO}$, and its sign rapidly changes to negative in the CO phase.
According to the Mott formula, the sign of the Seebeck coefficient of the DE system corresponds to the sign of the chemical potential $\mu$~\cite{Wei2009,Xing2009,Wang2011,Lundgren2014}.
Further, $\mu$ in $\alpha$-(BEDT-TTF)$_2$I$_3$ is always hole-like ($\mu < 0$) in the absence of carrier doping and interaction~\cite{Kobayashi2008, Tajima2012}. 
Thus, the positive sign of the Seebeck coefficient in the massless DE phase can be explained in the absence of carrier doping and interaction.
To our knowledge, however, the mechanism behind the sharp peak and the sign inversion of the Seebeck coefficient has not yet been elucidated. 

The theoretical derivation of the thermoelectric effect in condensed matter has attracted considerable attention. 
In DE systems, the Seebeck coefficient is an odd function of $\mu$ (bipolarity) forming positive and negative peaks. 
The magnitudes of the peaks are enhanced by the energy dependence of the relaxation time from impurity scattering in the massive DE~\cite{Sharapov},  and they are strongly affected by disorder and temperature~\cite{YamamotoFukuyama2}.
Recently, researchers have sought calculations ``beyond” the Mott formula, by incorporating the effects of electron correlation, impurity scattering, and phonon scattering on the Seebeck coefficient~\cite{JonsonMahan1990, Kontani, YamamotoFukuyama1, Ogata2017, Ogata2019}. 

In this study, we elucidate the effects of electron-hole asymmetry and Coulomb interactions on the Seebeck coefficient of $\alpha$-(BEDT-TTF)$_2$I$_3$ using an extended Hubbard model that  describes the electronic system of this material~\cite{Seo2000, Kobayashi2004, Kobayashi2011PRB, Matsuno2016, Omori2017, Ohki2018Crystals, Ohki2019} with mean-field approximation.
The Seebeck coefficient is calculated based on linear response theory for thermodynamic perturbations~\cite{Kubo, Luttinger, JonsonMahan1980, Kontani, Ogata2017}. The energy dependence of the relaxation time from impurity scattering is treated within the framework of the $T$-matrix approximation. 
We treat the chemical potential carefully, because the Seebeck coefficient is sensitive to it, and the temperature dependence of the chemical potential in the DE system is affected by electron-hole asymmetry, owing to the band structure and carrier doping~\cite{Kobayashi2008, Tajima2012}. 
In addition, the band structure is reshaped by the Coulomb interaction, which brings about a change in the temperature dependence of the chemical potential. 
Thus, the temperature dependence of the Seebeck coefficient in the DE system is strongly influenced by electron-hole asymmetry and the Coulomb interaction. 
These approaches allow us to understand the anomalous behavior of the Seebeck coefficient observed in the experiments~\cite{Kitamura, Konoike}. This behavior is the effect of drastic changes to the electronic state near the CO transition as a result of the Coulomb interaction.

The remainder of this paper is organized as follows. 
In Sec. II.A, we introduce an extended Hubbard model for describing $\alpha$-(BEDT-TTF)$_2$I$_3$ subject to a two-dimensional periodic boundary condition for calculating the electronic state. We formulate the Seebeck coefficient in Sec. II.B, based on the linear response theory of thermodynamic perturbations~\cite{Kubo, Luttinger, JonsonMahan1980, Kontani, Ogata2017}. 
The relaxation time in the $T$-matrix approximation is treated by the same framework used in previous research~\cite{Ruegg, Omori2017, Ohki2019}. 
In Sec. III.A, we show the temperature dependence of the chemical potential in the Hartree approximation. We compare the calculation results in the preceding study without Coulomb interaction, and consider the mechanism that produces this behavior based on the density of states and the given filling. 
In Sec. III.B, the numerical calculation results of the chemical potential dependence of the Seebeck coefficient are presented. 
Then, the filling is fixed to the value corresponding to the experiment, and we focus on the temperature dependence of the Seebeck coefficient in that case. 
In Sec. III.C, we discuss the contribution of the energy dependence of the relaxation time from impurity scattering to the temperature dependence of the Seebeck coefficient. 
The effects of electron doping are also discussed by comparing the results in a non-doping case. 
In Sec. III.D, we explain parameter tuning for electron-hole asymmetry, and we examine the changes to the Seebeck coefficient and Hall coefficient as this parameter changes, based on the Dirac cone model at several temperatures. 
We summarize our results in Sec. IV, and we position this study within recent work on the Seebeck coefficient in DE systems.

\section{Model and Formulation}

%
%
\subsection{Electronic states}
As a model that describes a pseudo-two-dimensional electronic system in $\alpha$-(BEDT-TTF)$_2$I$_3$, we use the two-dimensional (2D) extended Hubbard model~\cite{Seo2000}, where the effects of the insulating layer of I$^{3-}$ molecules are ignored ~\cite{Alemany}, except for their contribution to transfer integrals. 
The hopping energies up to the next nearest neighbor are obtained by a first-principles calculation~\cite{Kino}. 

%
\begin{figure}
\begin{centering}
\includegraphics[width=50mm]{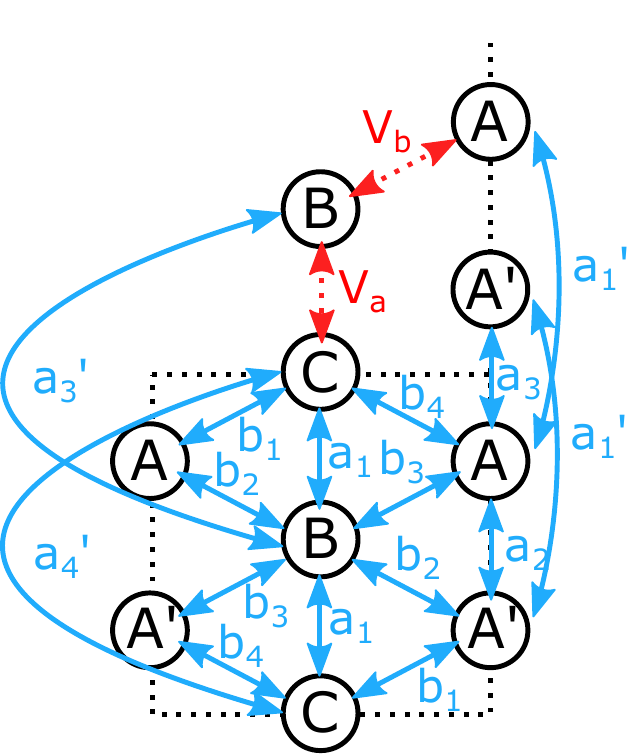}
\caption{\label{illustration}(Color online) Two-dimensional hopping network in the conduction plane of the organic conductor, $\alpha$-(BEDT-TTF)$_2$I$_3$. We consider up to the next-nearest-neighbor hopping energies, indicated by the solid arrows. Dashed arrows indicate the Coulomb interactions $V_a$ and $V_b$ between the nearest sites in the direction of the $a$ and $b$ axes.
}
\end{centering}
\end{figure}
Figure 1 shows a unit cell and a network of the hopping energies between each molecular site in the $a$--$b$ conduction plane. 
There are four sublattices, conventionally labeled A, A$'$, B, and C in the unit cell represented by the broken line. 
Here, inversion-symmetry points exist in the middle of the A and A$'$ sites, and at the B and C sites. 
As revealed by the analysis of Seo {\it et al.}~\cite{Seo2000}, the nearest-neighbor Coulomb interaction along the $a$ axis plays a principal role in driving the phase transition in $\alpha$-(BEDT-TTF)$_2$I$_3$ between the massless DE phase and the CO phase. 
Therefore, in addition to the on-site Coulomb interaction $U$, we only take into account the nearest-neighbor Coulomb interactions $V_a$ and $V_b$ indicated by the dashed arrow in Fig. 1. 
In what follows, lattice constants, the Boltzmann constant $k_B$, and the Planck constant $\hbar$ are taken as unity. Note that, throughout this paper, eV is used as the unit of energy.

The extended Hubbard model is given by
\begin{eqnarray}
H&=&{\sum_{<<i,{\alpha};j,\beta>>}}{\sum_{{\sigma}}t_{i,{\alpha};j,\beta}}a^{\dag}_{i,{\alpha},{\sigma}}a_{j,\beta,{\sigma}}+{\sum_{i,{\alpha}}}U n_{i,{\alpha},{\uparrow}}n_{i,{\alpha},{\downarrow}}\nonumber\\
&&+{\sum_{<i,{\alpha};j,\beta>}}{\sum_{{\sigma}{\sigma}'}}V_{i,{\alpha};j,\beta}n_{i,{\alpha},{\sigma}}n_{j,\beta,{\sigma}'}.
\end{eqnarray}
where $i$ and $j$ are the coordinates of the unit cell, $\alpha$ and $\beta$ represent the four sublattices ($=$ A, A$'$, B, and C) in the unit cell, and $\sigma$ is the spin index. 
Here, an electron number operator is defined by $n_{i,{\alpha},{\sigma}}=a^{\dag}_{i,{\alpha},{\sigma}}a_{i,{\alpha},{\sigma}}$. 
The first term is the kinetic energy, and the second term is the on-site Coulomb interaction.
The third term represents the nearest-neighbor Coulomb interaction, where $V_a$ is used for driving the CO transition, and we treat $V_b$ as a constant.
Further, $<\cdots>$ and $<<\cdots>>$ in the subscripts of summations refer to adding up the terms of the nearest and next-nearest neighbor, respectively.
$t_{i,{\alpha};j,\beta}$ shows the hopping between each site in Fig. 1 and is given as $t_{a1}=-0.0267$ ($-0.0101$), $t_{a2}=-0.0511$ ($-0.0476$), $t_{a3}=0.0323$ ($0.0093$), $t_{b1}=0.1241$ ($0.1081$), $t_{b2}=0.1296$ ($0.1109$), $t_{b3}=0.0513$ ($0.0551$), $t_{b4}=0.0152$ ($0.0151$), $t_{a1'}=0.0119$ ($0.0088$), $t_{a3'}=0.0046$ ($0.0019$), $t_{a4'}=0.0060$ ($0.0009$) at ambient pressure and temperature $T=0.0008$ ($0.03$: room temperature). 
In this study, we treat the temperature dependence of $t_{i,{\alpha};j,\beta}$ by a linear interpolation of the hopping values at $T = 0.0008$ and RT in Ref.~\cite{Kino}, as follows:
\begin{eqnarray}
&&t_{i,{\alpha};j,\beta}(T)=t_{i,{\alpha};j,\beta}({0.0008})\nonumber\\
&&{\hspace{0.5cm}}+\frac{t_{i,{\alpha};j,\beta}({\rm RT})-t_{i,{\alpha};j,\beta}({0.0008})}{0.0292}(T-0.0008).
\end{eqnarray}

By performing a Fourier inverse transform, $a_{i,\alpha,{\sigma}}={N_L}^{-{1}/{2}}\sum_{\bf k} a_{{\bf k}\alpha \sigma}e^{i{\bf k}\cdot{\bf r_i}}$ ($N_L$ is a system size), and Hartree approximation on Eq. (1), the Hamiltonian $H_{\rm MF}$ and its energy eigenvalue $E_{\nu \sigma}({\bf k})$ in the mean field approximation are obtained as follows:
\begin{eqnarray}
H_{\rm MF}&=&\sum_{{\bf k}}\sum_{\alpha \beta \sigma}\tilde{\epsilon}_{\alpha \beta \sigma}({\bf k})a^\dag_{{\bf k}\alpha \sigma}a_{{\bf k}\beta \sigma}- \sum_{\alpha}U_\alpha \langle n_{\alpha \uparrow}\rangle \langle n_{\alpha \downarrow}\rangle\nonumber\\
&& -\sum_{\alpha \beta \sigma \sigma'}V_{\alpha \beta} \langle n_{\alpha \sigma}\rangle \langle n_{\beta \sigma'}\rangle .
\end{eqnarray}
\begin{eqnarray}
\tilde{\epsilon}_{\alpha \beta \sigma}({\bf k})&=&\epsilon_{\alpha \beta}({\bf k})\nonumber\\
&&+\delta_{\alpha \beta}\left[ U_\alpha \langle n_{\alpha \bar{\sigma}}\rangle +\sum_{\beta' \sigma'}V_{\alpha \beta'}{\langle}n_{\beta' \sigma'}\rangle \right].
\end{eqnarray}
\begin{eqnarray}
E_{\nu \sigma}({\bf k})=\sum_{\alpha \beta}d^*_{\alpha \nu \sigma}({\bf k}) \tilde{\epsilon}_{\alpha \beta \sigma}({\bf k}) d_{\beta \nu \sigma}({\bf k})-\mu ,
\end{eqnarray}
where $\varepsilon_{\alpha \beta}({\bf k})=\sum_{\bm \delta}t_{\alpha \beta}e^{{\rm i}{\bf k}{\bm \cdot}{\bm \delta}}$ ($\bm \delta$ is a vector between unit cells), and $\nu =1, 2, 3, 4$ indicates a band index. 
Here, $d_{\alpha \nu \sigma}({\bf k})$ is a wave function diagonalizing $H_{\rm MF}$. 
The average electron number at each site is determined by $\langle n_{\alpha \sigma} \rangle = \sum_{{\bf k}, \nu}|d_{\alpha \nu \sigma}({\bf k})|^2f(E_{\nu \sigma}({\bf k}))$, where $f(E_{\nu \sigma}({\bf k}))=\left(1+\exp (E_{\nu \sigma}({\bf k})/T)\right)^{-1}$ is a Fermi distribution function, and the chemical potential $\mu$ is determined by the following equation:
\begin{equation}
\frac{3}{2}+\langle \delta n \rangle =\frac{1}{4}{\sum_{{\alpha}{\sigma}}}{\langle}n_{{\alpha}{\sigma}}{\rangle}.
\end{equation}
Because $\alpha$-(BEDT-TTF)$_2$I$_3$ has a $\frac{3}{4}$-filled band, the deviation of filling $\langle\delta n\rangle$ is zero when there is no impurity. 
We assume $\langle \delta n \rangle =10^{-6}$ ($1$ppm), because a small amount of electron doping has been confirmed in some samples of $\alpha$-(BEDT-TTF)$_2$I$_3$~\cite{Kobayashi2008,Tajima2012}.

A single particle green function $G^{0R}_{\alpha \beta}(\omega ,{\bf k})$ and the density of state $\mathcal{N}(\omega)$ are given by 
\begin{eqnarray}
G^{0R}_{\alpha \beta}(\omega ,{\bf k})&=&\sum_{\nu \sigma}\frac{d^*_{\alpha \nu \sigma}({\bf k})d_{\beta \nu \sigma}({\bf k})}{\hbar\omega-E_{\nu \sigma}({\bf k})+i\eta},\\
\mathcal{N}(\omega)&=&-\pi^{-1}{\rm Im}\left[ {\rm Tr}\hspace{0.1cm} G^{0R}(\omega) \right].
\end{eqnarray}

The critical value of $V_a$ for the CO transition at $T=0$ is $V_a^C = 0.198$. 
In the following, we compare numerical results in three cases: $(U, V_a ,V_b )=(0, 0, 0)$ (non-interacting case); $(U, V_a ,V_b )=(0.4, 0, 18, 0.05)$ (massless DE phase appearing at any temperatures); and $(U, V_a ,V_b )=(0.4, 0.199, 0.05)$ (CO transition occurring at $T_{\rm CO}=0.002$).

\subsection{Transport property}

The Seebeck coefficient is given by the Nakano--Kubo formula for linear response theory~\cite{Kubo, Luttinger, JonsonMahan1980, Kontani, Ogata2017}. 
The Seebeck coefficient at a low temperature limit $S(\mu ,T\sim0)$ is calculated using the Mott formula: 
\begin{equation}
S(\mu ,T\sim0)=-\frac{\pi^2}{3e}T\left[\frac{\partial}{\partial \mu}\ln{\sigma(\omega ,T=0)}\right]_{\hbar\omega=\mu}
\end{equation}
where $e > 0$ is the elementary charge. 

Moreover, $S(\mu, T)$ at finite temperatures~\cite{Ogata2019,YamamotoFukuyama1} is given by 
\begin{eqnarray}
S(\mu ,T)&=&\frac{L_{12}}{L_{11}}\\
L_{11}&=&{\mathscr L}^{(0)}_y=\sigma_{yy}\\
L_{12}&=&-\frac{1}{eT}{\mathscr L}^{(1)}_y
\end{eqnarray}
where $L_{11}$ and $L_{12}$ are coefficients for the electric field $\bf E$, and the temperature gradient $-{\bm \nabla} T$ of the current density $\bf j$ and heat flow density ${\bf j}^Q$ is defined by 
\begin{eqnarray}
{\bf j}&=& L_{11}{\bf E} + L_{12}(-{\bm \nabla} T),\\
{\bf j}^Q&=&L_{21}{\bf E} + L_{22}(-{\bm \nabla} T),
\end{eqnarray}
where $L_{11}$ is equal to the DC conductivity. 
In this study, $L_{12}$ and the direct current conductivity $L_{11}=\sigma_{yy}$ in the direction of the $a(y)$ axis of the conduction plane are calculated using the expression of the transport coefficient ${\mathscr L}^{(m)}_{y}$, as follows: 
\begin{eqnarray}
{\mathscr L}^{(m)}_{y}&=&\int d\omega \left(-\frac{df}{d\omega} \right)(\hbar \omega)^{m}\Phi_y(\omega),\\
\Phi_y(\omega)&=&\frac{4 e^2}{N_L}\sum_{{\bf k} \nu}\left|{\bf v}^y_\nu({\bf k})\right|^2\tau_\nu(\omega,{\bf k})\delta(\hbar \omega-E_\nu({\bf k})), 
\end{eqnarray}
where $N_L$ indicates the system size and the velocity matrix ${\bf v}^y_\nu({\bf k})$ is a derivative of the energy eigenvalue 
$\tilde{\epsilon}_{\alpha \beta \sigma}({\bf k})$ regarding the wave number $k_y$. This is obtained by converting it to a band representation, ${\bf v}^y_{\nu \nu' \sigma}({\bf k})=\sum_{\alpha \beta}d^*_{\alpha \nu \sigma}({\bf k})v^y_{\alpha \beta \sigma}({\bf k})d_{\beta \nu' \sigma}({\bf k})$, using the wave function $d_{\alpha \nu \sigma}({\bf k})$. 

Regarding the effect of impurity scattering on the Seebeck coefficient~\cite{JonsonMahan1990}, the impurity potential term is derived as follows: 
\begin{eqnarray}
H_{\rm imp}=\frac{V_0}{N_L}\sum_{{\bf k} {\bf q} \sigma} \sum_{i \alpha}^{N_{\rm imp}} e^{-i{\bf q}\cdot {\bf r}_i}a^{\dag}_{{\bf k}+{\bf q} \alpha \sigma}a_{{\bf k} \alpha \sigma}, 
\end{eqnarray}
and is added as a perturbation to $H_{\rm MF}$. 
Here $\sum_{i \alpha}^{N_{\rm imp}}$ means the summation over all impurities in the system. 
${\bf r_i}$ ($i = 1, \cdots , N_{\rm imp}$) represents a coordinate about unit cells, and $N_{\rm imp}$ is the total number of impurities. 
$H_{\rm imp}$ is treated within the $T$-matrix approximation to include the energy dependence with the relaxation time $\tau_\nu(\omega)$~\cite{Shon, Proskurin, Ruegg, Omori2014, Omori2017, Ohki2019}. 
As a result, the retarded self-energy $\Sigma^R_{\nu \sigma}(\omega,{\bf k})$ and the damping constant $\gamma_{\nu \sigma}(\omega,{\bf k})$ are obtained as follows: 
\begin{eqnarray}
\Sigma^R_{\nu \sigma}(\omega,{\bf k})&=&c_{\rm imp}\sum_{\alpha}\frac{V_0\left|d_{\alpha \nu \sigma}({\bf k}) \right|^2}{1-\frac{V_0}{N_L}\sum_{{\bf k}'}G^{0R}_{\alpha \sigma}(\omega,{\bf k}')},\\
\gamma_{\nu \sigma}(\omega,{\bf k})&=&\frac{\hbar}{2\tau_{\nu \sigma}(\omega,{\bf k})}=-{\rm Im}\Sigma^R_{\nu \sigma}(\omega,{\bf k})\nonumber\\
&=&c_{\rm imp}\sum_{\alpha}\frac{ \left|d_{\alpha \nu \sigma}({\bf k})\right|^2\left\{ \pi V_0^2 \mathcal{N}_{\alpha \sigma}(\omega)\right\}}{1+\left\{ \pi V_0 \mathcal{N}_{\alpha \sigma}(\omega)\right\}^2}, 
\end{eqnarray}
where the impurity density $c_{\rm imp}=\frac{N_{\rm imp}}{N_L}=0.02$ and the strength of the impurity potential $V_0=0.1$. 
We assume that the impurities are distributed uniformly. 
As the above equation indicates, the relaxation time within the $T$-matrix approximation $\tau_\nu(\omega)$ is inversely proportional to $c_{\rm imp}$ and shows the energy dependence that reflects the shape of the density of state $\mathcal{N}(\omega)$. 
More specifically, $\tau_\nu(\omega)$ diverges when $c_{\rm imp}$ or $\mathcal{N}(\omega)$ become zero. 
In order to avoid the divergence of $\tau_\nu(\omega)$, we set the cutoff to $5\times10^6$ $\hbar$(eV)$^{-1}$ (1 $\hbar$(eV)$^{-1} \sim 6.58\times10^{-16}$ s) caused by the effects of scattering beyond the $T$-matrix approximation.

\section{Numerical Results}

%
%
\subsection{Effect of Coulomb interaction on electronic states}
%
%

%
%
\begin{figure}
\begin{centering}
\includegraphics[width=60mm]{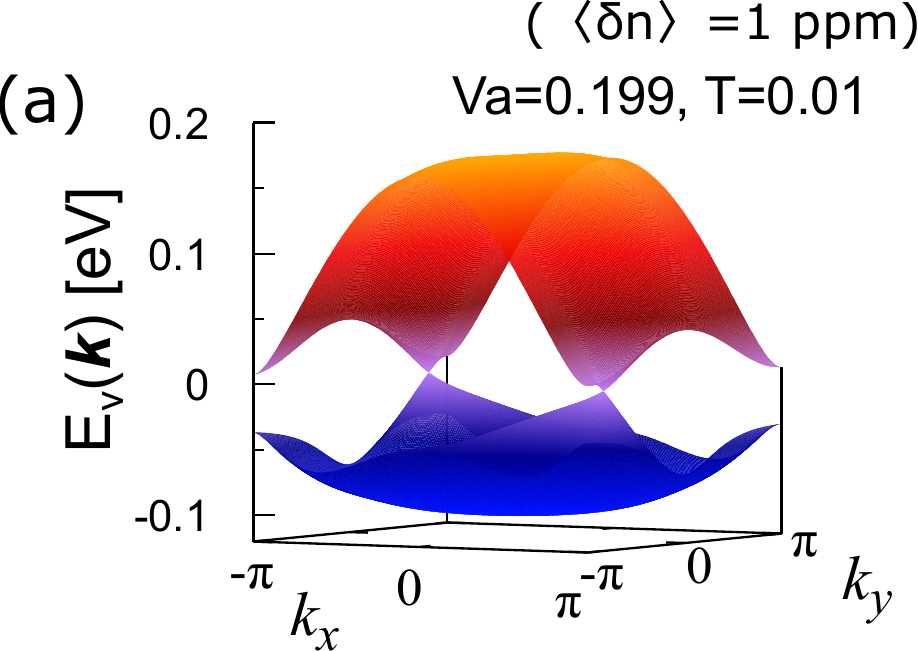}
\includegraphics[width=65mm]{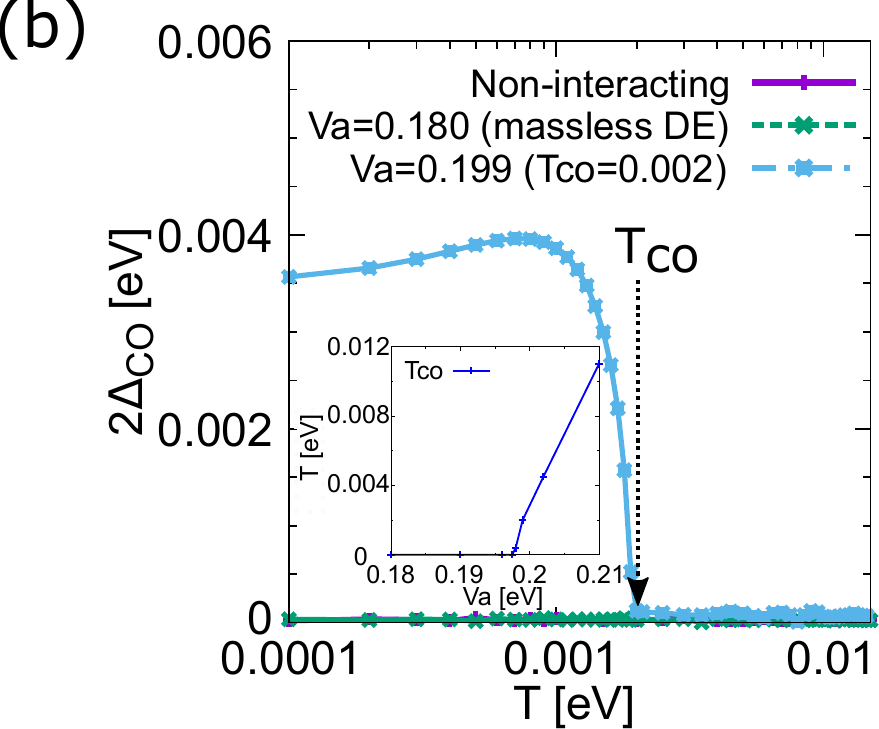}
\includegraphics[width=65mm]{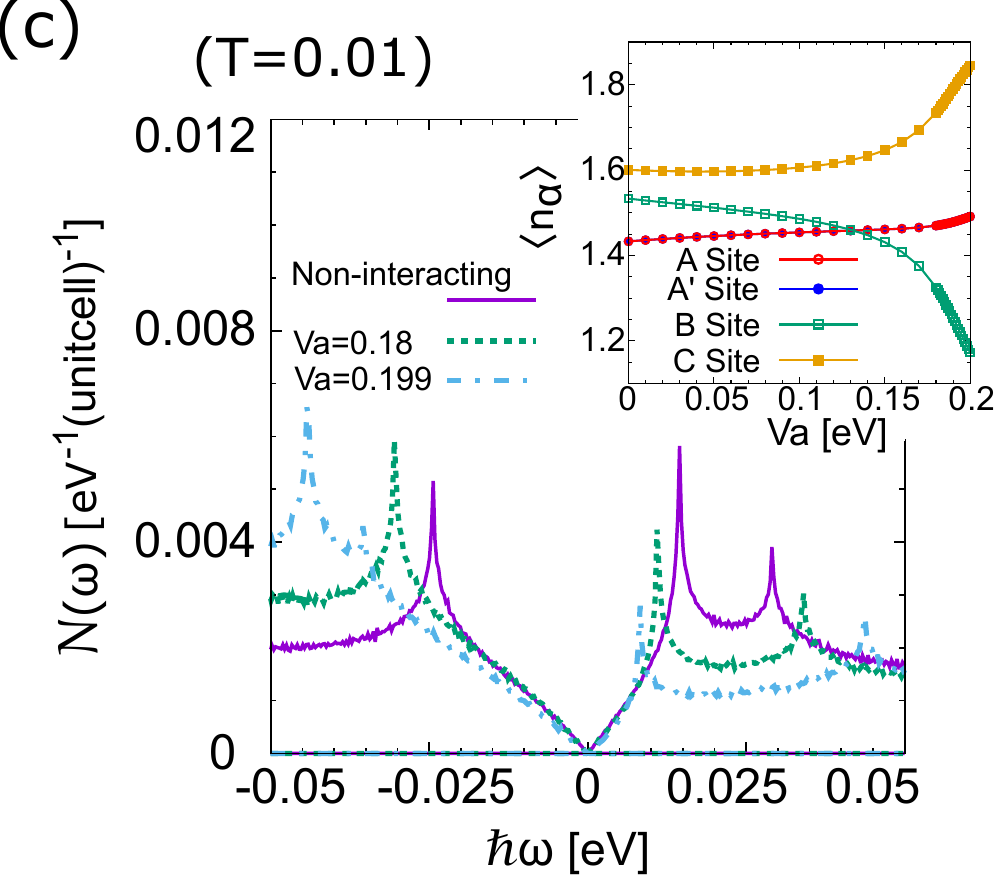}
\includegraphics[width=65mm]{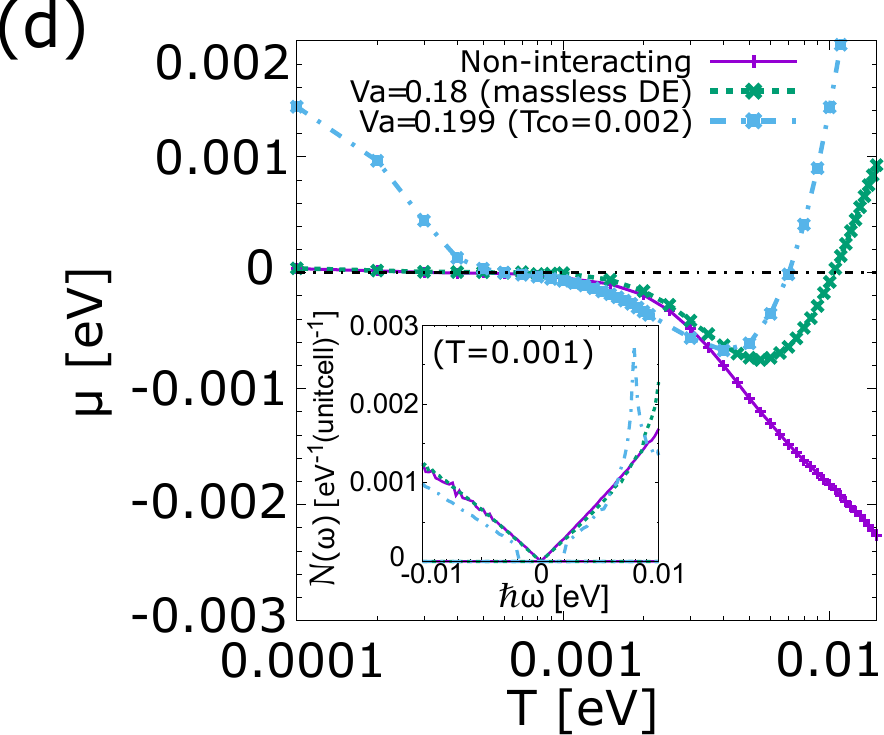}
\caption{\label{illustration}(Color online) (a) Energy eigenvalue $E_\nu({\bf k})$ with $\nu=1$ (red band) and $2$ (blue band) near the chemical potential at  $(U, V_a ,V_b )=(0.4, 0.199, 0.05)$ and $T = 0.01$. (b) $T$ dependence of the energy gap $2\Delta_{\rm CO}$. (c) Density of state $\mathcal{N}(\omega)$ at $T = 0.01$. (d) $T$ dependence of $\mu$ in the non-interacting case (solid line), $V_a = 0.18$ (dashed line), and $V_a = 0.199$ (single-dotted chain line). $\mu$ is measured from the contact point for $2\Delta_{\rm CO} = 0$, and from the center of energy gap for $2\Delta_{\rm CO} \neq 0$.
The inset of (b) shows the $V_a$ dependence of $T_{\rm CO}$. 
The $V_a$ dependence of a charge density of each sublattice in the unit cell is plotted in the inset of (c). 
The density of state $\mathcal{N}(\omega)$ at $T = 0.001$ is also shown in the inset of (d). 
}
\end{centering}
\end{figure}
%
%
Figure 2(a) shows the energy bands near the chemical potential at $V_a = 0.199$ and $T = 0.01$. 
There is a pair of tilted Dirac cones, and the Dirac points are located in the vicinity of the chemical potential.
We show the $T$ dependence of the CO gap $2\Delta_{\rm CO}$ in the non-interacting case, $V_a = 0.180$, and $V_a = 0.199$ in Fig. 2(b). 
$2\Delta_{\rm CO}$ is determined as the indirect gap between $E_1({\bf k})$ and $E_2({\bf k})$, and becomes a finite value below $T_{\rm CO} =0.002$. 
The inset of Fig. 2(b) shows the $V_a$ dependence of $T_{\rm CO}$, where the inversion symmetry is broken in the CO phase. 
The CO gap shows the non-monotonic temperature dependence at low temperatures, owing to the temperature dependence of the transfer integrals. 
Figure 2(c) shows the density of states $\mathcal{N}(\omega)$ near the Fermi energy at the massless DE phase at those three $V_a$ values. 
Because the Hartree term induced by $V_a$ enhances the electron-hole asymmetry in the energy bands, the density of states in the band $\nu =1 (2)$ increases (decreases) near the Fermi energy as $V_a$ increases. 
Such a deformation of the energy band is caused by the relative change to the charge density at each sublattice with the increase of $V_a$, as shown in the inset of Fig. 2(c). 
Figure 2(d) shows the $T$ dependence of $\mu$ measured from the contact point or the center of energy gap at those three $V_a$ values. 
In the non-interacting case, $\mu$ decreases monotonically as $T$ increases, and $\mu$ is negative (hole-like) except at very low temperatures $T < 2 \times 10^{-4}$, because the Van Hove singularity in the band $\nu =1$ is closer to the Dirac point than that of the band $\nu =2$~\cite{Kobayashi2008}. 
At very low temperatures $T < 2 \times 10^{-4}$, $\mu$ is positive, owing to the small amount of electron doping $\langle \delta n \rangle =10^{-6}$.
In cases where $V_a = 0.180$ and $V_a = 0.199$, $\mu$ becomes positive at high temperatures near $T \sim 0.01$, because the electron-hole asymmetry of the density of states in the energy scale of $T$ is enhanced by the Hartree term, as shown in Fig. 2(c). 
In the case where $V_a = 0.199$, $\mu$ has a large positive value near $T=0$, because $2\Delta_{\rm CO}$ is finite below $T_{\rm CO} =0.002$, as shown in Fig. 2(b) and the inset of Fig. 2(d). 
Thus, the $T$ dependence of $\mu$ is strongly influenced by electron-hole asymmetry and the Coulomb interaction.

\subsection{Temperature dependence of the Seebeck coefficient}
%
%
%

%
\begin{figure}
\begin{centering}
\includegraphics[width=70mm]{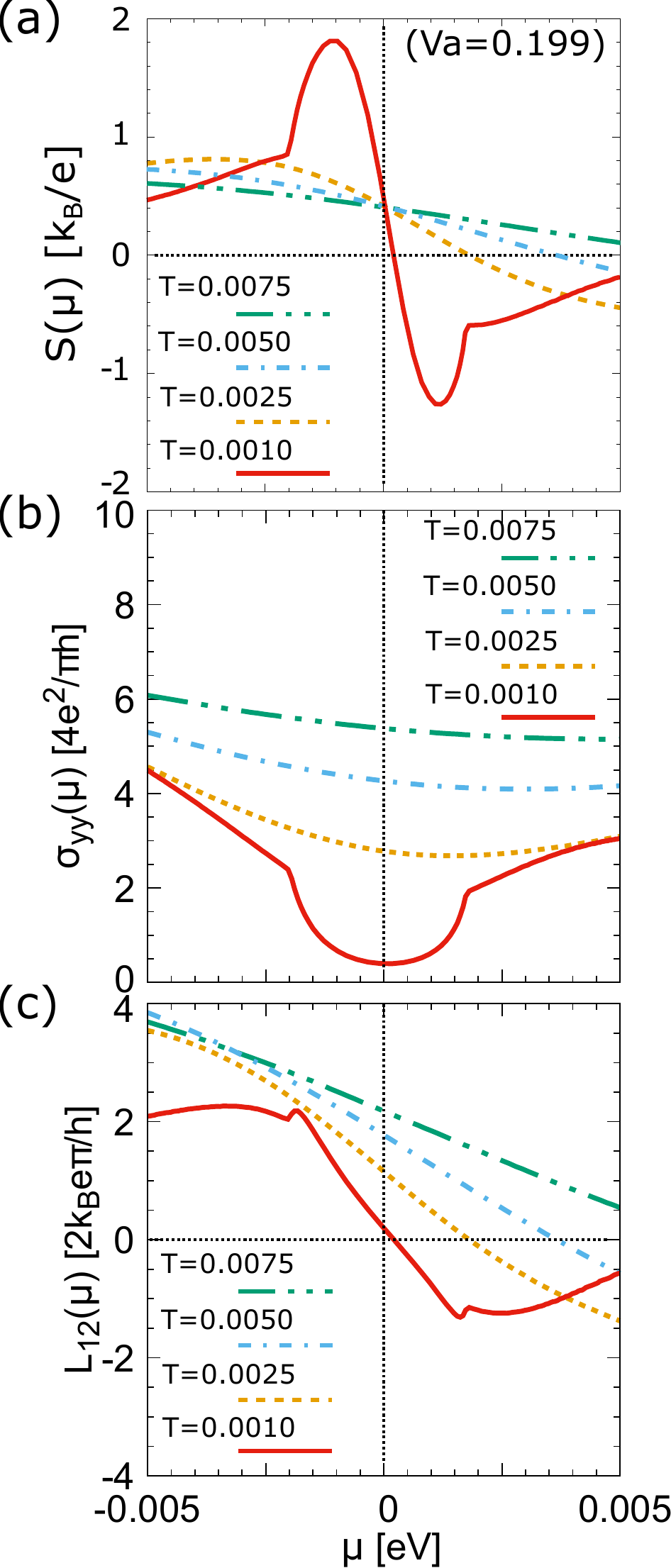}
\caption{\label{illustration}(Color online) Chemical potential $\mu$ dependence of (a) the Seebeck coefficient $S$ in units of $k_B/e \simeq 10^2\mu V / K$, (b) the DC conductivity $L_{11}=\sigma_{yy}$ in units of the universal conductivity $4e^2/\pi h$, and (c) $L_{12}$ in units of $2k_Be\pi/h$ at $V_a=0.199$ for temperature $T = 0.0075$ (double-dotted chain line), $0.005$ (single-dotted chain line), $0.0025$ (dashed line), and $0.001$ (solid line). Here, $\mu$ is varied by the calculation performed with changing $\langle \delta n \rangle$ in Eq. (6). 
}
\end{centering}
\end{figure}
The chemical potential $\mu$ dependence of the Seebeck coefficient $S$, the DC conductivity $L_{11}=\sigma_{yy}$ in the denominator of $S$, and $L_{12}$ in the numerator of the $S$ are shown in Fig. 3(a), (b), and (c) at several temperatures for $V_a = 0.199$. 
As temperature $T$ decreases, $S(\mu)$ forms gentle positive and negative peaks which come from the function shape of $L_{12}(\mu)$~\cite{Sharapov, YamamotoFukuyama2}. 
On the other hand, $S(\mu)$ shows specifically large positive and negative peaks in $\left|\mu\right| < \Delta_{\rm CO}\simeq 0.002$ at $T$ = $0.001$ ($<T_{\rm CO} =0.002$). 
These large peaks in $|\mu| < \Delta_{\rm CO}$ and discontinuous changes near $\mu \simeq \pm\Delta_{\rm CO}$ arises mainly from the sharp decrease of $\sigma_{yy}(\mu)$ at $|\mu| < \Delta_{\rm CO}$ as $T$ decreases (See Fig. 3(b) and (c)). 
We note that, overall, $S(\mu =0)>0$ and $S(\mu)$ shift to positive values. 
Because the Seebeck coefficient is influenced by the carrier doping $\langle \delta n \rangle =10^{-6}$ as well as the electron-hole asymmetry of the band structure, as discussed in Subsection III.D, the sign of $S(\mu)$ does not have a one-to-one correspondence with the sign of $\mu = \mu(\langle \delta n \rangle, T)$. 
In the following, $\langle \delta n \rangle$ is fixed as $10^{-6}$ unless otherwise noted. 
%
%

\begin{figure}
\begin{centering}
\includegraphics[width=80mm]{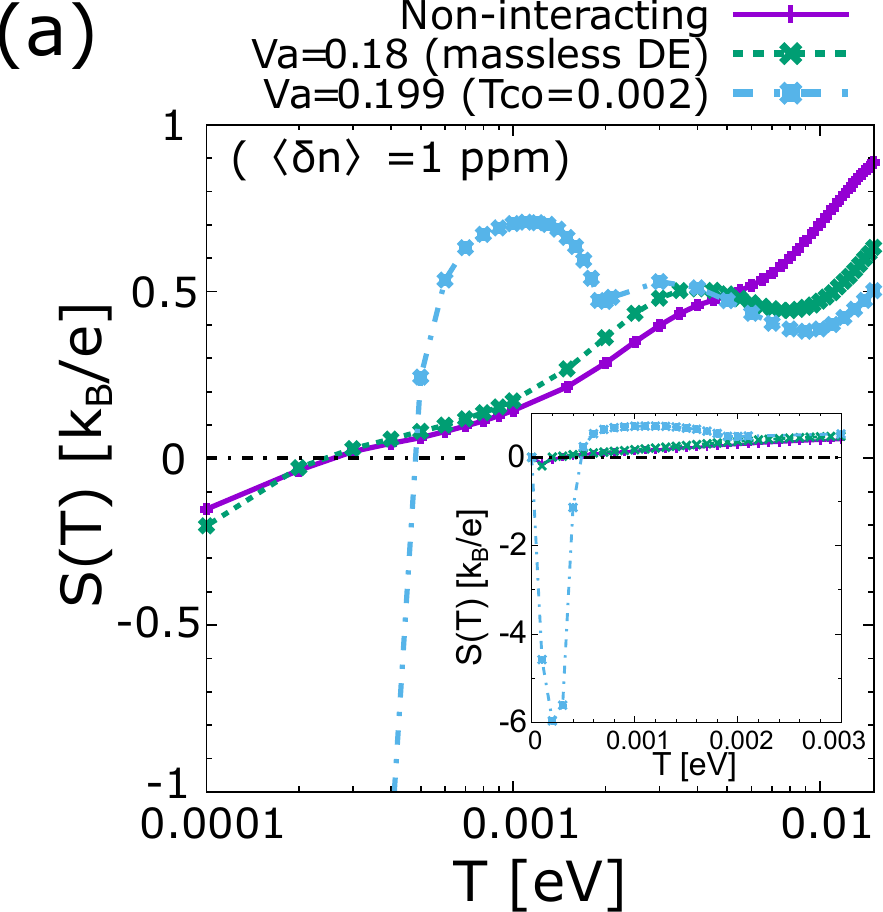}
\includegraphics[width=80mm]{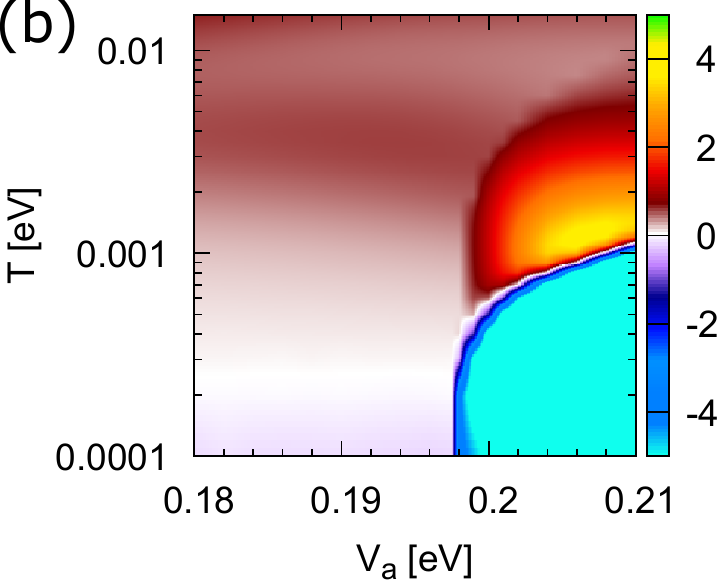}
\caption{\label{illustration}(Color online) (a) Temperature dependence of Seebeck coefficient $S$ at $\langle\delta n\rangle=1$ ppm (electron-doped) for the non-interacting case (solid line), $V_a=0.18$ (dashed line), and $V_a=0.199$ (single-dotted chain line). Inset: Temperature dependence of $S$ near $T = 0$ for those cases. $S$ at $T = 0$ is calculated by the Mott formula. (b) Color plot of the Seebeck coefficient $S$ at $\langle\delta n\rangle=1$ ppm (electron-doped) versus $V_a$ and $T$. 
}
\end{centering}
\end{figure}
Figure 4(a) shows the temperature dependence of $S$ in the non-interacting case, $V_a = 0.180$, and $V_a = 0.199$.
Here, $S(T)$ in the non-interacting case decreases monotonously as the temperature decreases, and changes the sign from positive to negative at temperature $T = 2 \times 10^{-4}$, corresponding to the sign change of $\mu$ from negative (hole-like) to positive (electron-like), as shown in Fig. 2(d). 
As $V_a$ increases, $S(T)$ near $T\sim0.01$ decreases, because $\mu$ near $T\sim0.01$ increases and becomes positive, as shown in Fig. 2(d).
As a result, a gentle peak is induced by $V_a$ around $T=0.005$.
This gentle peak is similar to that observed in Ref.~\cite{Kitamura, Konoike}. 
At $V_a=0.199$, we find a sharp peak with $S(T)$ just below $T_{\rm CO}$, as a result of a sudden decrease in $L_{11}$ and the energy dependence of the relaxation time with impurity scattering, as discussed in the Subsection III.C.
Moreover in $T<T_{\rm CO}$, because $\mu$ suddenly changes its sign from negative to positive owing to the existence of a finite $2\Delta_{\rm CO}$ (see Figs. 2(b) and 2(d)), $S(T)$ rapidly decreases and changes its sign from positive to negative, as shown by the single-dotted line in Fig. 4(a). 
This behavior qualitatively demonstrates the peak structure observed near $T_{\rm CO}$ in experiments. 
The inset of Fig. 4(a) shows $S(T)$ at the low temperature region ($0\le T\le 0.003$). 
Here, $S(T)$ has a negative value at low temperatures, owing to the slight electron doping. 
At the limit of $T\to0$, $S$ becomes zero according to the Mott formula, but if the temperature is slightly finite, the contribution of $S(T)\to0$ from $T\sim 0$ competes for the contribution, and $S(T)$ remains finite on account of carrier doping. Thus, $S(T)$ changes its value considerably. 
Figure 4(b) shows a color plot of the Seebeck coefficient: $S$ versus $V_a$ and $T$. 
The temperature at which point the sign of $S$ inverts at $T < T_{\rm CO}$ shifts to a higher temperature as $V_a$ increases (Note that we only calculated a few points to plot Fig. 4(b) and the oscillatory behavior near the phase transition in this figure is an error on the plot caused by the lack of data points). 
\subsection{Effect of the energy dependence of the relaxation time and electron doping}
%
%
%
\begin{figure*}[tb]
\begin{centering}
\includegraphics[width=130mm]{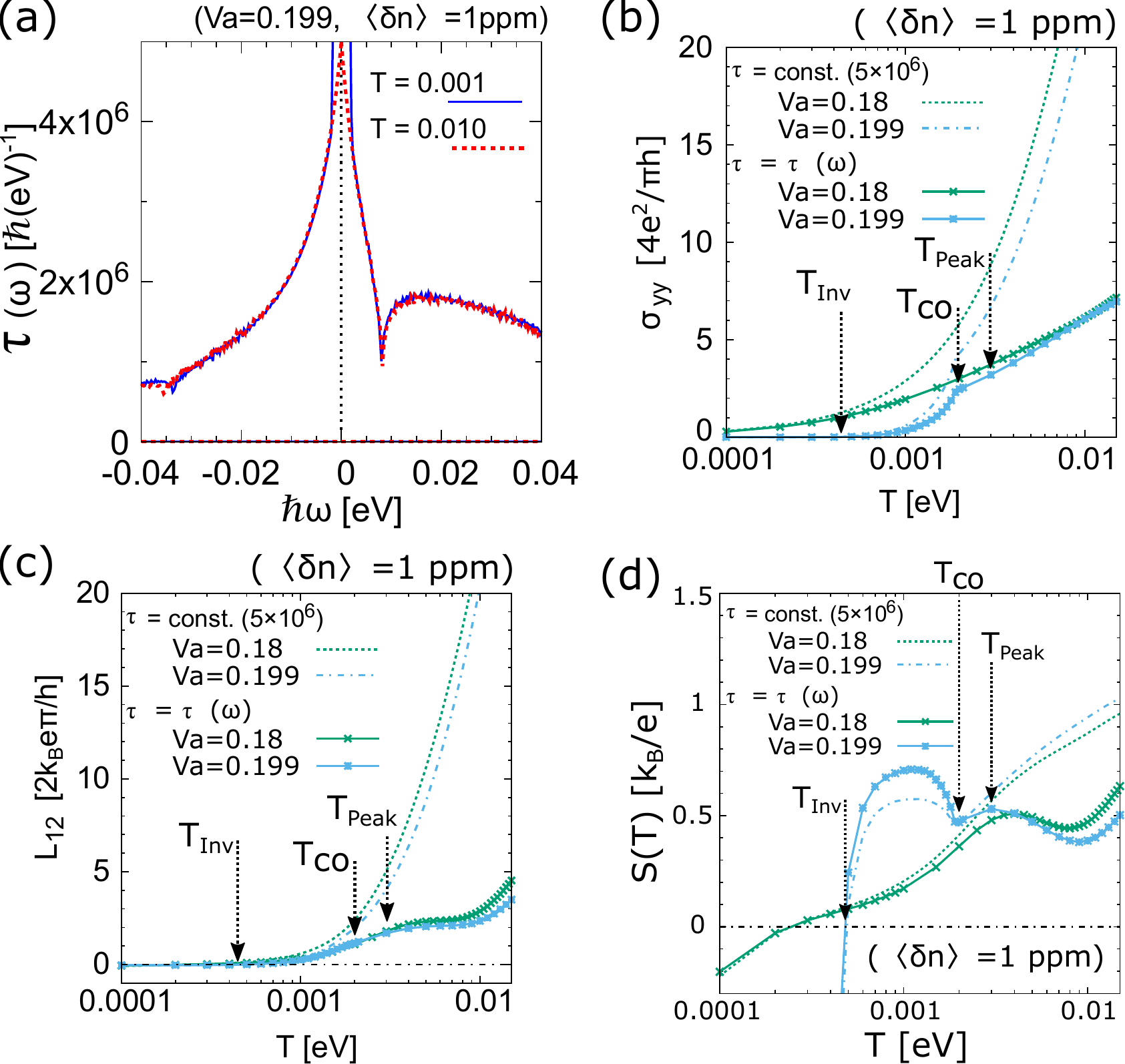}
\caption{\label{illustration}(Color online) (a) Energy dependence of the relaxation time $\tau(\omega)$ in units of $\hbar$(eV)$^{-1}$ at $|{\bf k}|=k_F$ in the case of $V_a=0.199$, plotted for $T=0.01$ (dashed line), and $T=0.001$ (solid line). 
(b), (c), and (d) Temperature dependence of DC conductivity $L_{11}=\sigma_{yy}$ in units of the universal conductivity $4e^2/\pi h$, $L_{12}$ in units of $2k_Be\pi/h$, and Seebeck coefficient $S$ in units of $k_B/e$ at $\langle\delta n\rangle=1$ ppm (electron-doped) for $V_a=0.18$ and $V_a=0.199$ in the case of $\tau(\omega)$ in the $T$-matrix approximation  (solid line with point) and in the case of a constant $\tau =5 \times 10^6$ (others). 
}
\end{centering}
\end{figure*}
In this subsection, we focus on the effect of impurity scattering and the contribution of the energy dependence of the relaxation time $\tau(\omega)$ on the $T$ dependence of the Seebeck coefficient $S(T)$. 
Figure 5(a) shows the $\omega$ dependence of $\tau(\omega)$ at the wave number $|{\bf k}|=k_F$ and $V_a = 0.199$ considering impurity scattering according to the $T$-matrix approximation with $c_{\rm imp}=0.02$ and $V_0 =0.1$. 
As shown in the Fig. 5(a), $\tau (\omega)$ is about inversely proportional to the density of state $\mathcal{N}_{\alpha \sigma}(\omega)$ and reflects the shape of $\mathcal{N}_{\alpha \sigma}(\omega)$ at each temperature (e.g., the Van Hove singularity, Dirac point, and energy gap, regarding which see Fig. 2(c)). 
$T_{\rm Peak}$ is defined as the temperature where the peak structure appears in the massless DE phase, and $T_{\rm Inv}$ is characterized by the sign inversion of $S(T)$ in $T < T_{\rm CO}$ for visualization purposes.
Figures 5(b) and 5(c) show the temperature dependence of DC conductivity $\sigma_{yy}(T)$ and $L_{12}(T)$ corresponding to the denominator and numerator of $S(T)$ shown in Fig. 5(d), respectively,  in cases with $\tau(\omega)$ (solid line with point) and a constant $\tau =5 \times 10^6$. 
There are drastic differences between these cases.
We found that the gentle peak structure at $T_{\rm Peak}$ in the massless DE phase is derived from $L_{12}(T)$ with $\tau(\omega)$. 
The sudden increase in the absolute value of $S(T)$ at $T < T_{\rm CO}$, however,  is caused by the decrease of $\sigma_{yy}(T)$.
The sign inversion temperature $T_{\rm Inv}$ of  $S(T)$ corresponds to that of $L_{12}(T)$, and it is determined by both the $T$ dependence of $\mu (T)$ (Fig. 2(d)) and the $\mu$ dependence of $S(\mu)$ (Fig. 3)
The thorn-like structure between $T_{\rm Inv}$ and $T_{\rm CO}$ appears only when the sample is slightly electron-doped, $\langle \delta n \rangle > 0$. 

%
%
\begin{figure}
\begin{centering}
\includegraphics[width=75mm]{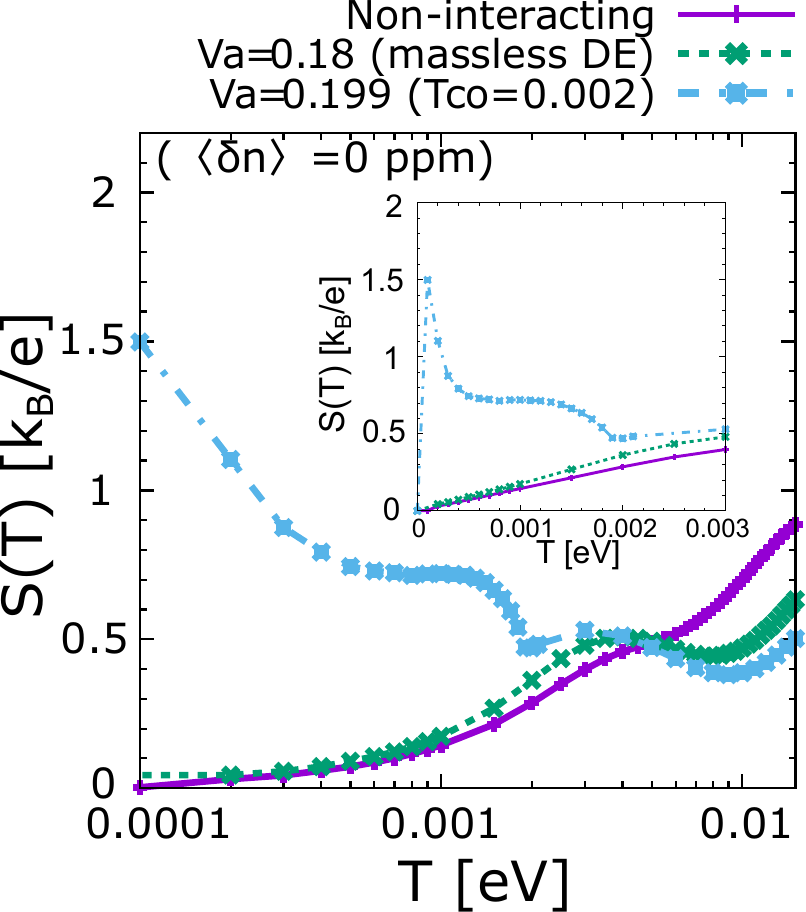}
\caption{\label{illustration}(Color online) Temperature dependence of the Seebeck coefficient $S$ in the case where $\langle\delta n\rangle=0$ ppm (non-doping case) for the non-interacting case (solid line), $V_a=0.18$ (dashed line), and $V_a=0.199$ (single-dotted chain line). 
The inset shows the temperature dependence of $S$ in the linear scale near $T = 0$ in the above three interaction values ($0\le T\le 0.003$). 
}
\end{centering}
\end{figure}
Figure 6 shows $S(T)$ in a non-doping case ($\langle \delta n \rangle = 0$ ppm). 
In this case, because the chemical potential does not reverse its sign from negative to positive at low temperatures ($T<0.001$), $S(T)$ is always positive, as shown in Fig. 6. 
At $T <T_{\rm CO}$, $S(T)$ increases suddenly and has a large positive value at low temperatures, because $\sigma_{yy}(T)$ reaches zero, although $S(T)$ becomes zero at $T = 0$, as shown in the inset of Fig. 6.

\subsection{{Change to electron-hole asymmetry and Seebeck and Hall coefficients}}
Next, we consider the relationship between the electron-hole asymmetry of the energy band and the Seebeck coefficient using a tilted Weyl model~\cite{Kobayashi2008} to represent the tilted Dirac cone of $\alpha$-(BEDT-TTF)$_2$I$_3$. 
In general, when the Seebeck coefficient is calculated using a symmetrical electron-hole energy band, the sign inversions of the carrier and Seebeck coefficient correspond to each other. 
However in the previous subsection, the Seebeck coefficient $S(T)$ showed a positive value at high temperatures, even though the chemical potential $\mu$ was positive (Fig. 2(d) and Fig. 4(a)). 
Moreover, the positive chemical potential at finite temperatures, from the contribution of the Hartree term ($\mu>0$ at $T\sim0.01$), does not agree with the temperature dependence of the Hall coefficient, as observed in experiments with $\alpha$-(BEDT-TTF)$_2$I$_3$~\cite{Tajima2012}. 
These results can never be obtained with calculations using a symmetrical electron-hole band, indicating that $S>0$ ($S<0$) when the carrier is hole-like (electron-like). 
Therefore, in this subsection, we show that the sign of the Seebeck or Hall coefficients calculated with an asymmetrical electron-hole energy band does not always match the sign of the carrier, and the energy where their signs invert shifts from the effects of the electron-hole asymmetry.

We introduce a tilted Weyl Hamiltonian that represents the low-energy band dispersion of $\alpha$-(BEDT-TTF)$_2$I$_3$, as follows~\cite{Kobayashi2008,Tissermond}:
\begin{eqnarray}
H=\sum_{\rho=x,y,z,0}\tilde{\bf k}\cdot{\bf v}_\rho({\bf k}'_0)\sigma_\rho-\mu + \frac{\hbar^2 k^2}{2m}{\rm X}. 
\end{eqnarray}
In the first term of Eq. (20), $\sigma_0$ means a unit matrix and $\sigma_x$, $\sigma_y$, $\sigma_z$ indicate the Pauli matrices. 
${\bf k'}_0$ is a wave number which indicates infinitesimally different from the Dirac point ${\bf k}_0$, and ${\bf \tilde{k}} = {\bf k}-{\bf k'}_0$ is defined as a wave number measured from ${\bf k'}_0$. 
Also, ${\bf v}_\rho({\bf k_0'})$ is calculated by the velocity matrix $u_{\nu,\nu'}^{\tau}({\bf k})$ defined as follows: 
\begin{equation}
u_{\nu\nu'}^{\tau}({\bf k})=\sum_{\alpha\beta}d^*_{\alpha\nu}({\bf k})\frac{\partial \tilde{\epsilon}_{\alpha \beta}({\bf k})}{\partial k_{\tau}}d_{\beta\nu'}({\bf k}),
\end{equation}
where $\tau = x, y$ and $\tilde{\epsilon}_{\alpha \beta}({\bf k})$ and $d_{\alpha \nu}({\bf k})$ are given by Eqs. (4) and (5). 
Each component of ${\bf v}_\rho({\bf k_0'})$ are respectively given by ${\bf v}_x = {\rm Re}[{\bf u}_{12}({\bf k'}_0)]$, ${\bf v}_y = -{\rm Im}[{\bf u}_{12}({\bf k'}_0)]$, ${\bf v}_z = \frac{1}{2}\left[{\bf u}_{11}({\bf k'}_0)-{\bf u}_{22}({\bf k'}_0)\right]$, and ${\bf v}_0 = \frac{1}{2}\left[{\bf u}_{11}({\bf k'}_0)+{\bf u}_{22}({\bf k'}_0)\right]$~\cite{Kobayashi2008}. 
The second term of the Hamiltonian $H$ is a chemical potential term which only shifts the origin of energy, and the third term is distorting the Dirac cone and changes the electron-hole asymmetry of the energy dispersion~\cite{Tissermond}. 
This curvature term is derived from the second derivative of $\tilde{\epsilon}_{\alpha \beta}({\bf k})$ about the wave number $k_\tau$ \cite{Kobayashi2011PRB} by assuming the isotropy on the differential of $\tilde{\epsilon}_{\alpha \beta}({\bf k})$ about each $k_\tau$. 
Here, we control the sign and magnitude of the curvature term using mass change ratio $X$ which changes in the range of $-1 < X <1$ and a mass parameter $m$ is set as a constant ($m=1$). 
Eq. (20) leads to the next energy dispersion: 
\begin{eqnarray}
E^{\pm}_{\bf k}&=&{\bf \tilde{k}}\cdot{\bf v}_0({\bf k'}_0)\pm\sqrt{\sum_{\nu=x,y,z}\left[{\bf \tilde{k}}\cdot{\bf v}_\nu({\bf k'}_0)\right]^2}\nonumber\\
&&-\mu + \frac{\hbar^2 k^2}{2m}{\rm X}
\end{eqnarray}
As an example, the density of states at $X = -1$ and $X=1$ are shown in Figure 7(a).

To obtain the Seebeck coefficient and the Hall coefficient, $L_{11}=\sigma_{yy}$ and $L_{12}$ are calculated using the transport coefficient ${\mathscr L}^{(m)}_{y}$ (Eq.~(15)) with the energy dispersion (Eq.~(22)). 
Here, the Hall conductivity is calculated by the following approximated formula, exclusively considering the intra-band contribution~\cite{Fukuyama19691, Fukuyama19692, Kobayashi2008}:
\begin{eqnarray}
\sigma_{xy} &=& \frac{4e^3H}{3\pi c}\sum_{\rho=\pm}\int \int dk_x dk_y \int dE \nonumber\\
&&\times f'(\epsilon)\left[ \left( \frac{\partial E^\rho_{\bf k}}{\partial k_x} \right)^2\frac{\partial^2 E^\rho_{\bf k}}{\partial k_y^2}-\frac{\partial E^\rho_{\bf k}}{\partial k_x}\frac{\partial E^\rho_{\bf k}}{\partial k_y}\frac{\partial^2 E^\rho_{\bf k}}{\partial k_x \partial k_y} \right]\nonumber\\
&&\times\frac{\Gamma^3}{\left[(E-E^\rho_{\bf k}+\mu(T))^2+\Gamma^2 \right]^3}
\end{eqnarray}
where $H$ is a magnetic field and $\Gamma$ is a phenomenologically introduced damping constant for impurity scattering. 
Here, $\Gamma$ depends on the temperature, such that $\Gamma=\Gamma_0+\theta T$. We set $\Gamma_0=10^{-5}$ and $\theta=10^{-3}$. 
The DC conductivity along the $b(x)$ axis $\sigma_{xx}$ is also calculated using the same formula as $\sigma_{yy}$, and the Hall coefficient $R_H$ is obtained by
\begin{eqnarray}
R_H = \frac{\sigma_{xy}}{H\sigma_{xx}^2}. 
\end{eqnarray}
In this study, we assume electronic carriers, and we set the chemical potential to $\mu = 0.0001$.
%
%
\begin{figure}
\begin{centering}
\includegraphics[width=90mm]{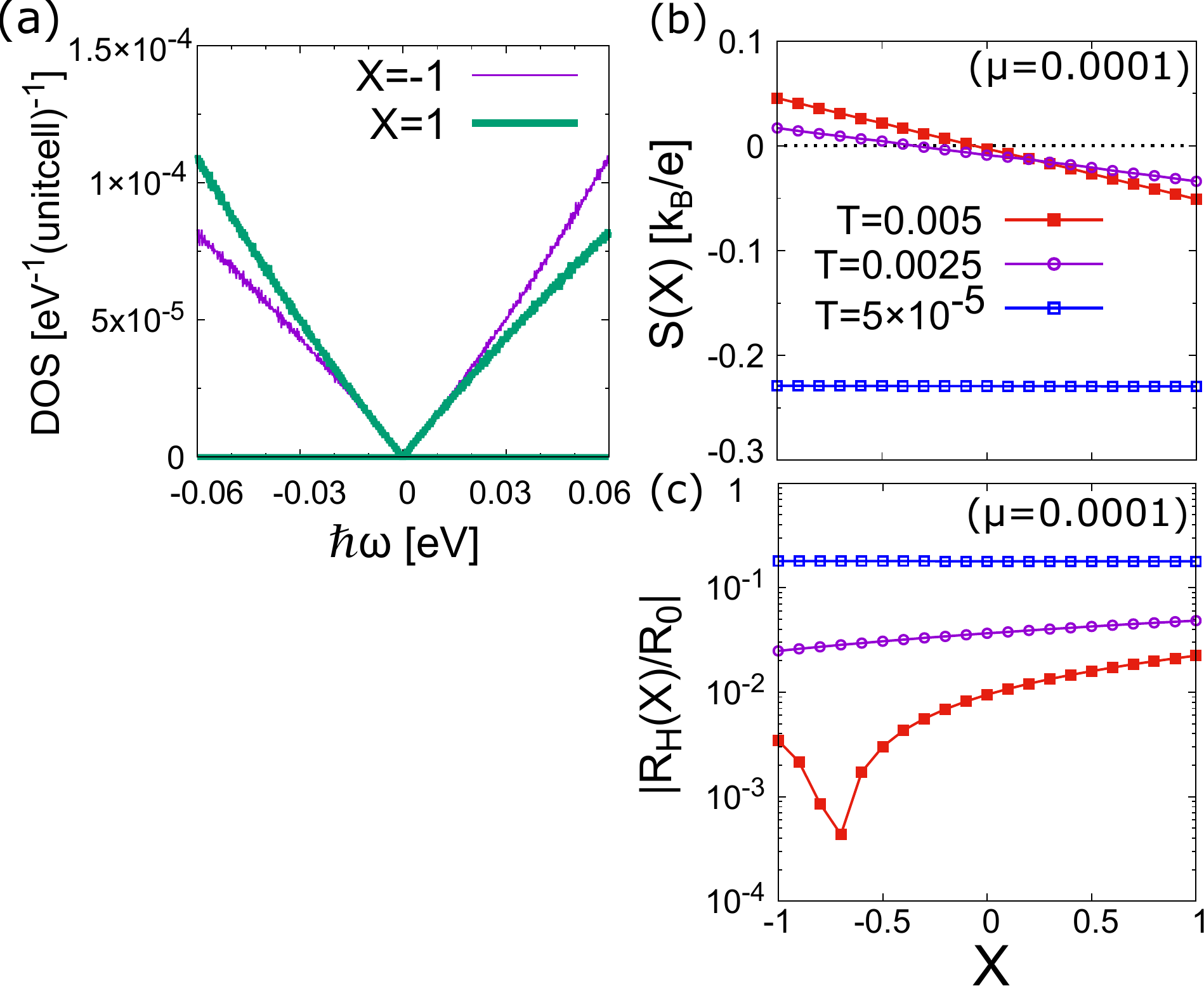}
\caption{\label{illustration}(Color online) (a) Density of the state at the mass change rate $X=-1$ (thin line) and $X=1$ (thick line). The $X$ dependence of (b) the Seebeck coefficient $S$ and (c) the absolute value of the Hall coefficient $|R_H(X)/R_0|$, where $T=0.005$, $0.0025$, and $5 \times 10^{-5}$. We assume an electronic band structure, and we set the chemical potential to $\mu=0.0001$.
}
\end{centering}
\end{figure}
%
%
Figure 7(b) shows the Seebeck coefficient with respect to the mass change ratio $X$ in $\mu=0.0001$ and the three temperature cases: $T=0.005$, $0.0025$, and $5\times10^{-5}$. 
In the case of $T=5\times10^{-5}$, which is the lowest temperature among the three cases, the Seebeck coefficient is independent of $X$ and becomes a negative constant, reflecting the positive $\mu$. 
However, as the temperature increases with $T = 0.0025$ and $0.005$, $S(X)$ gradually behaves proportionally to $X$, and a range of $X$ appears such that $S(X)$ is positive. 
The sign of $S$ is determined by the sign of $L_{12}$, as shown in Eq.~(10). 
A reason for this $T$- and $X$-dependent $S$ behavior is perhaps that the change in the electron-hole asymmetry more easily affects the value of $L_{12}$ as the temperature increases. 
Because the higher energy part of the density of states more positively contributes to the value of $L_{12}$, the density of states is reflected by change in electron-hole asymmetry. 

By contrast, the absolute value of the Hall coefficient $|R_H(X)/R_0|$ with respect to the mass change ratio $X$ is shown in Figure 7(c) for $\mu=0.0001$ where $T=0.005$, $0.0025$, and $5\times10^{-5}$. 
Here, we set $R_0={\pi^2v_x^2}/{ec\Gamma_0^2}$ and $v_x=|{\bf v}_x|\sim0.01$. 
The Hall coefficient $R_H$ also reflects the shape of the density of state as it reaches higher temperatures, and a range of $X$ appears such that $R_H$ is positive, despite $\mu>0$. (The sharp ``V"-shaped structure of $| R_H (X) / R_0 |$ in Fig. 7 refers to the sign inversion of $R_H$.) 
%
%
%

\section{Summary and Discussion}

In this study, we investigated the effects of the electron correlation and the electron-hole asymmetry of the energy band on the Seebeck coefficient with an extended Hubbard model that describes the DE system of the organic conductor $\alpha$-(BEDT-TTF)$_2$I$_3$.
We found that they affect the Seebeck coefficient through the energy dependence of the relaxation time from impurity scattering.
As a result, the Seebeck coefficient has a gentle peak near $T=50$ K, in contrast to cases when we ignore the electron correlation effect or when using a constant relaxation time.
Furthermore, we found that a thorn-like structure of the Seebeck coefficient appears just above the CO phase transition temperature, which can be explained in two steps: 
1) The sudden decrease in conductivity that accompanies the phase transition causes an abrupt increase in the absolute value of the Seebeck coefficient. 
2) Assuming slight electron doping, the Seebeck coefficient drops sharply and inverts its sign as a result of the drastic sign change of the chemical potential, owing to the emergence of an energy gap. 
This behavior in massless DE and CO phases qualitatively agrees with the experimental results~\cite{Kitamura, Konoike}.

We also showed that the signs of the Seebeck and Hall coefficients do not necessarily correspond to the sign of the chemical potential, owing to the effect of electron-hole asymmetry.
We found that by distorting the band dispersion in the Weyl model, the Seebeck coefficient at finite temperature becomes insensitive to changes in the chemical potential, although it reflects the shape of the energy band. 
Thus, the Seebeck and Hall coefficients at finite temperatures show different $\mu$ dependence from those at $T=0$.

Finally, the nearest-neighbor Coulomb interaction $V_a$ was used as a control parameter for the CO transition, rather than the actual pressure dependence, and we used transfer integrals at ambient pressure. 
The temperature dependence of the Coulomb interaction, which was ignored in this time, also needs attention naturally when $V_a$ plays a significant role in the phase transition. 
Furthermore, we only treated the elastic scattering by impurities and the Seebeck coefficient was calculated using the Mott formula. 
However, the inelastic scattering by electron--electron and the electron--phonon which contribute to the behavior of the Seebeck coefficient~\cite{Kontani, Ogata2019} can not be ignored in finite temperature. 
It is known that the electron correlation effects play important roles in $\alpha$-(BEDT-TTF)$_2$I$_3$~\cite{KinoFukuyama, Seo2000,Tanaka2016,Ishikawa2016, Hirata2016, Matsuno2017, Hirata2017, Matsuno2018}. 
Phonon drag may also contribute to the peak structure near $T=0.005$ of the Seebeck coefficient, although electron--phonon scattering was ignored in this study. 
In future research, we should calculate considering these effects respectively and explore difference from this study, and aim to reproduce the temperature dependence of the Seebeck coefficient shown in experiments more accurately.

\begin{acknowledgments}
The authors would like to thank T. Yamamoto, H. Fukuyama, S. Onari, and H. Kontani for fruitful discussions, and Y. Yamakawa for advice on the numerical calculations. 
This work was supported by MEXT (JP) JSPJ (JP) (Grants No. 19J20677, No. 19K03725, No. 19H01846, and No. 15K05166). 
\end{acknowledgments}




\end{document}